\documentclass[lettersize,journal]{IEEEtran}
\usepackage{amsmath,amsfonts}

\usepackage{array}
\usepackage[caption=false,font=normalsize,labelfont=sf,textfont=sf]{subfig}
\usepackage{stfloats}
\usepackage{url}
\usepackage{verbatim}
\usepackage{graphicx}
\usepackage{subcaption}
\usepackage{booktabs}

\usepackage{algpseudocode}
\usepackage{algorithm}

\def\BibTeX{{\rm B\kern-.05em{\sc i\kern-.025em b}\kern-.08em
    T\kern-.1667em\lower.7ex\hbox{E}\kern-.125emX}}
\usepackage{balance}
\begin{document}
\title{How to Use the IEEEtran \LaTeX \ Templates}
\author{IEEE Publication Technology Department
\thanks{Manuscript created October, 2020; This work was developed by the IEEE Publication Technology Department. This work is distributed under the \LaTeX \ Project Public License (LPPL) ( http://www.latex-project.org/ ) version 1.3. A copy of the LPPL, version 1.3, is included in the base \LaTeX \ documentation of all distributions of \LaTeX \ released 2003/12/01 or later. The opinions expressed here are entirely that of the author. No warranty is expressed or implied. User assumes all risk.}}

\markboth{ }%
{How to Use the IEEEtran \LaTeX \ Templates}

    \title{Near-Threshold Voltage Massive MIMO Computing  }

\author{
\IEEEauthorblockN{Mikael Rinkinen\IEEEauthorrefmark{1}, Mehdi Safarpour\IEEEauthorrefmark{2}, Shahriar Shahabuddin\IEEEauthorrefmark{3}, Olli Silvén\IEEEauthorrefmark{2}, Lauri Koskinen\IEEEauthorrefmark{1}}\\
\IEEEauthorblockA{\IEEEauthorrefmark{1}Circuit and Systems Unit, University of Oulu, Finland}\\
\IEEEauthorblockA{\IEEEauthorrefmark{2}Center for Machine Vision and Signal Analysis, University of Oulu, Finland\\
\IEEEauthorblockA{\IEEEauthorrefmark{3} Oklahoma State University, USA\\
\{firstname.lastname\}@oulu.fi} \\
\{firstname.lastname\}@okstate.edu}
}


\maketitle

\vspace{-10pt}
\begin{abstract}
 Massive MIMO systems have the potential to significantly enhance spectral efficiency, yet their widespread integration is hindered by the high power consumption of the underlying computations. This paper explores the applicability and effectiveness of Algorithm-Based Fault Tolerance (ABFT) for massive MIMO signal processing to tackle the reliability challenge of Near Threshold Computing (NTC).

 We propose modifying matrix arithmetic Newton iteration MIMO algorithm  to seamlessly integrate ABFT to detect any computational errors by inspecting the final result. The overhead from ABFT depends largely on the matrix dimensions, which in this context are dictated by the number of user equipments involved in the computation. 

NTC is a promising strategy for reducing the energy consumption in digital circuits by operating transistors at extremely reduced voltages. However, NTC is highly susceptible to variations in Process, Voltage, and Temperature (PVT) which can lead to increased error rates in computations. Traditional techniques for enabling NTC, such as dynamic voltage and frequency scaling guided by circuit level timing error detection methods, introduce considerable hardware complexity and are difficult to implement at high clock frequencies.  In this context ABFT has emerged as a lightweight error detection method tailored for matrix operations without requiring any modifications on circuit-level and can be implemented purely in software.

To evaluate the system across a range of supply voltages, the MIMO accelerator was implemented on a reconfigurable hardware platform. Experimental results demonstrate that for sufficiently large problem sizes, the proposed method achieves a 36\% power saving compared to baseline, with only an average of ~3\% computational overhead, at default clock frequency. The proposed approach is hardware-agnostic and can be integrated without requiring any modifications to the circuit or extensive cell-level characterization. These results indicate that combining ABFT with near-threshold operation provides a viable path toward energy-efficient and robust massive MIMO processors.
\end{abstract}

\begin{IEEEkeywords}
 Low power, baseband processing, 5G.
\end{IEEEkeywords}

\IEEEpeerreviewmaketitle

\section{Introduction}

Massive Multiple-Input Multiple-Output (MIMO) systems have emerged as a transformative technology to improve spectral efficiency in wireless communications~\cite{huai2014low}. This has led to their adoption in modern communication standards such as IEEE 802.11n, IEEE 802.11e, and LTE~\cite{huai2014low}.

By enabling simultaneous transmission over multiple spatial streams, MIMO systems maximize the use of the available radio spectrum. However, this gain in throughput comes at the cost of increased computational burden on the digital baseband processor. As the number of transmit antennas and supported User Equipment (UE) scales—reaching up to 128, 256, or even 512 antennas and dozens of users, the complexity of signal detection and channel processing grows rapidly~\cite{studer2007matrix}. Although several low-complexity or nonlinear detection algorithms have been proposed to mitigate this, they still impose considerable computational demands~\cite{huai2014low}. This translates directly into higher power consumption, increased hardware cost, and more stringent thermal requirements—challenges that are especially critical in components such as Remote Radio Heads (RRHs)~\cite{bursalioglu2016rrh}.

\begin{figure}[t!]
  \begin{center}
  \includegraphics[width=3in]{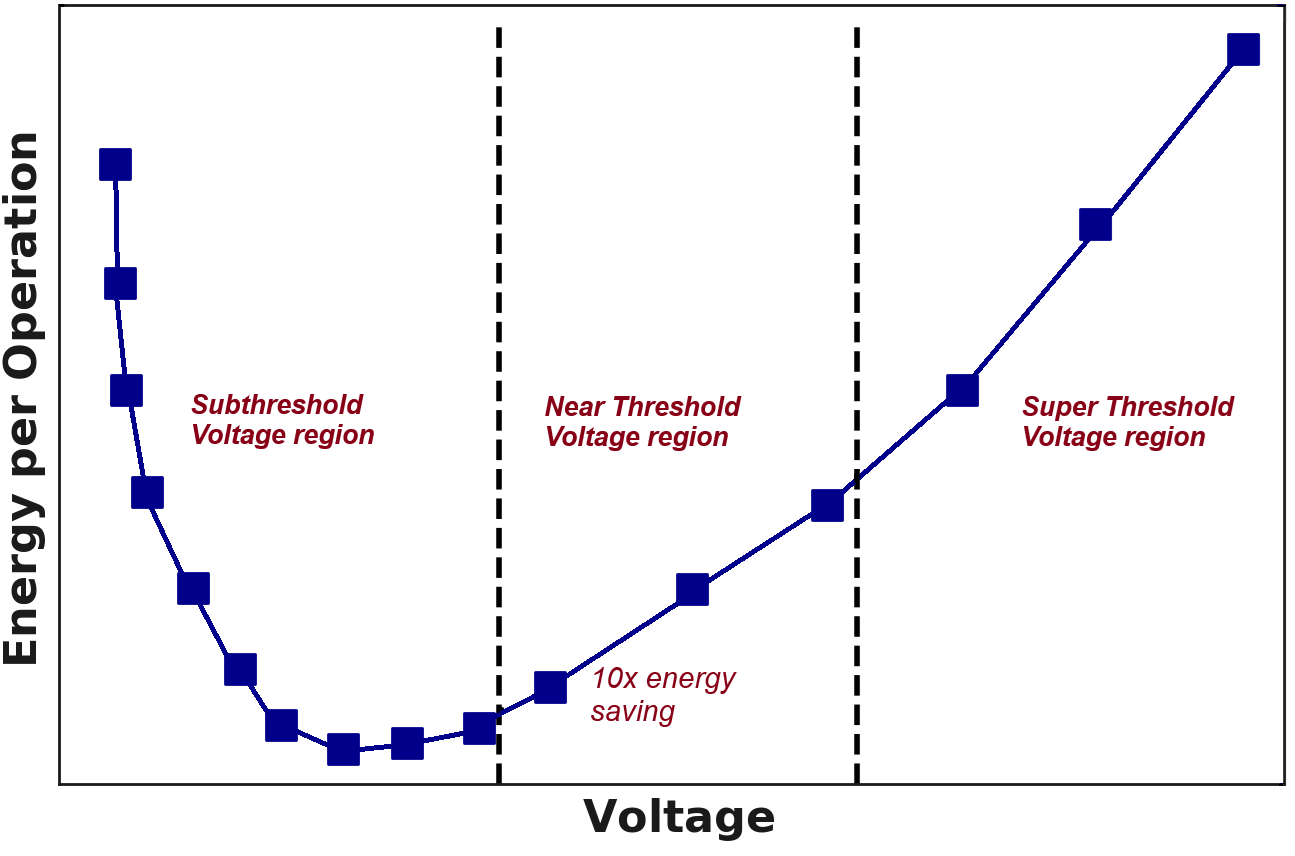}
  \vspace{-5pt}
  \caption{Lowering the supply voltage can greatly improve energy efficiency in processors~(adapted based on results of \cite{jain2012280mv}), but operating close to the threshold voltage increases the likelihood of timing errors.}
  \label{fig:NearThreshold}
  \end{center}
  \vspace{-20pt}
\end{figure}

Voltage scaling is a widely used strategy to reduce energy consumption in digital processors. Because dynamic power scales quadratically and static power scales linearly with voltage, lowering the supply voltage can yield substantial energy savings. Nevertheless, commercial chips are typically operated with conservative voltage margins to guard against Process, Voltage, and Temperature (PVT) variations~\cite{Koutsovasilis2020Impact, krishna2022global}. These guardbands, while ensuring reliability across diverse operating conditions, can lead to up to 30\% excess power consumption~\cite{Koutsovasilis2020Impact}. On top of this, reducing voltage beyond voltage guardband down to vicinity of threshold voltage of transistors, i.e., near-threshold and subthreshold regions, promises 10x to 100x improvement in energy efficiency improvements~\cite{uytterhoeven2021design}, as illustrated in Fig.~\ref{fig:NearThreshold}.

A major challenge in voltage scaling is the timing errors due exacerbated sensitivity to PVT variations. While several circuit-level error detection and correction methods exist, they are often too complex and introduce large overheads. Furthermore, those are incompatible with off-the-shelf hardware and closed-source IPs~\cite{wang2005180, do2023evaluating}.

To address this, prior work~\cite{safarpour2021algorithm} introduced a software-level solution by integrating Algorithm-Based Fault Tolerance (ABFT)~\cite{huang1984algorithm} as the error detection mechanism instead of circuit based timing error detection methods. ABFT enables data-level error detection during computation, much like how Error Correction Codes (ECC) protects stored or transmitted data. Unlike hardware approaches, ABFT requires no circuit changes or detailed characterizations.

This is particularly interesting for MIMO systems, which rely heavily on matrix operations for detection, such as Gram matrix computation and linear system solving. These operations are natural candidates for ABFT protection. Leveraging ABFT in this context enables reliable operation at reduced voltages, allowing significant power savings without compromising correctness. This work presents a modified MIMO algorithm that seemlessly integrates ABFT and presents evaluation of this concept on undervolted FPGA device, quantifying the power savings and computational trade-offs.

\section{Background}

\subsection{Error detection for reduced voltage operation}

Hardware-based strategies involve making adjustments to circuits during the design phase to estimate timing or catch timing errors in real-time \cite{gundi2021effort}. One straightforward method to monitor timing errors is through slack estimation techniques. These methods incorporate logic delay chains that simulate the longest delay paths in the original design. They act as early warning systems, detecting errors before the main delay path does when voltage is reduced. However, challenges such as local temperature variations, process differences, circuit aging, and noise introduce a significant trade-off between energy conservation and reliability \cite{krishna2022global}.

Another approach, offline calibration \cite{zhao2017robust}, determines the minimum voltage a device needs through intermittent offline measurements and updates a calibration table. While commercially adopted, this method's energy efficiency contribution is limited due to its susceptibility to changes in the operating environment and circuit aging \cite{jiang2022fodm}.

 \begin{figure}[t!]
  \begin{center}
  \includegraphics[width=3.2in]{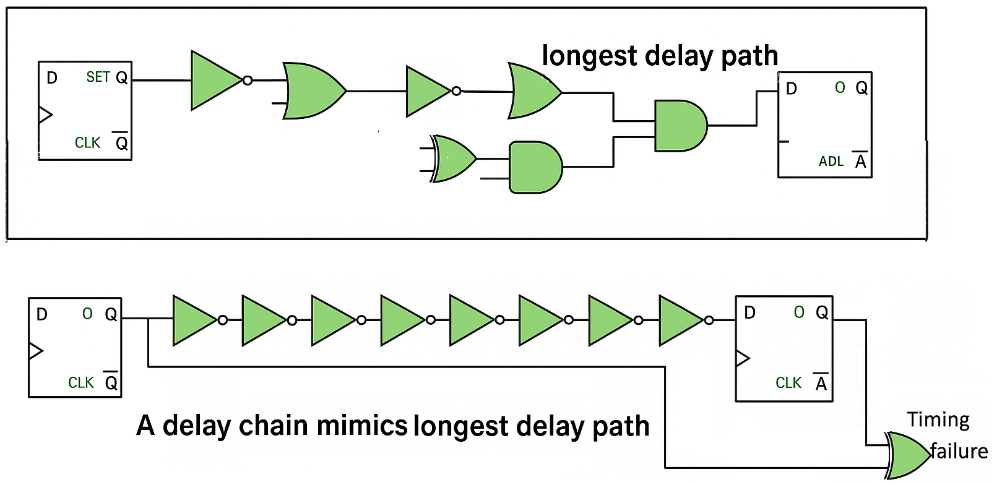}
  \vspace{-5pt}
  \caption{The bottom circuit is a delay chain of gates minimizing timing behavior of critical path in the top circuit. The solution acts as slack timing sensor to detect possible timing errors from voltage down-scaling. }\label{circuit_error_detection}
  \end{center}
  \vspace{-10pt}
\end{figure}

Instead of replicating the longest delay paths, i.e., Fig.~\ref{circuit_error_detection}, that is susceptible to PVT variations, timing errors can be identified within the delay path itself using Timing Error Detection (TED) systems. These systems utilize Error Detection Sequence (EDS) circuits, where a secondary register, working alongside the primary one, checks the output of the logic path within a pipeline stage. The secondary register samples with a slightly delayed clock. A difference in results between the primary and secondary registers indicates a late arrival signal and a timing error \cite{maragos2021pvt}. Wang et al. \cite{wang2003energy} employed a TED system in their FFT to enable operation at low voltage. However, the adoption of slack estimation and TED systems for low voltage is hindered by design and testing complexities. Recently, Jiang et al. \cite{jiang2022fodm} introduced automated tools to facilitate FPGAs in benefiting from TED. Notably, few commercially successful designs based on TED have entered the market, likely due to high design costs \cite{nunez2021energy}.

Algorithmic alternatives such as replication schemes, e.g., Dual Modular Redundancy (DMR), incur excessive overhead costs, undermining the benefits derived from voltage down-scaling and rendering them impractical. While Error Correction Codes have been successfully applied to voltage scaling for memories \cite{yalcin2014exploiting}, they are not suitable for detecting errors in computing. However, there exist algorithmic, i.e., coding-based, approaches that offer an interesting avenue to explore as alternatives to hardware-based error detection methods. Algorithm-Based Fault Tolerance (ABFT) introduced by Huang and Abraham \cite{huang1984algorithm} specifically  can detect and to some degree correct computing errors of matrix operations \cite{filippas2021low,safarpour2021algorithm,rinkinen2024shavette,aliagha2025scissors}. ABFT has proven to be robust and appealing, characterized by an overhead ratio of $O(1/N)$, where $N$ denotes the matrix size. Similarly for Fast Fourier Transform (FFT) the Parseval's identity property has been proposed as low cost robust computing error detection solution \cite{safarpour2022lofft}, e.g., for OFDM based scheme. There even exist algorithmic schemes for detecting errors in convolution operations of Deep Neural Networks \cite{marty2020safe,rinkinen2024shavette,safarpour2021low} and decoders \cite{valkama2023low}. Those were previously exploited for enabling robust and energy efficient low voltage operation \cite{safarpour2021high}.

 \begin{figure}[t!]
  \begin{center}
  \includegraphics[width=3.4in]{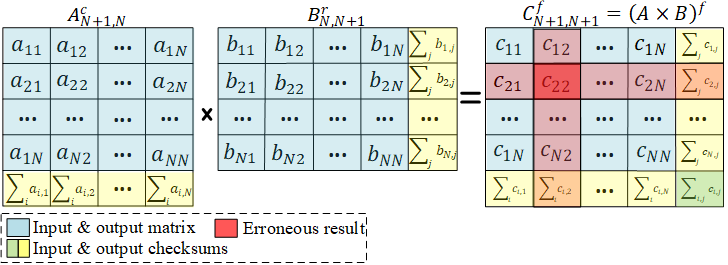}
  \vspace{-5pt}
  \caption{By adding checksum row/column to the input matrices, the computational errors in output matrix can be detected according to ABFT scheme (adapted from \cite{safarpour2021algorithm}). }\label{fig:overheads}
  \end{center}
  \vspace{-10pt}
\end{figure}

As mentioned earlier, the ABFT schemes come with trade offs with respect to matrix size and hence this paper tried to find the optimum configuration to maximize gains from low voltage operation that is enabled by leveraging ABFT.

\subsection{MIMO Detection}
MIMO detection is one of the most computationally intensive tasks for a wireless receiver. A MIMO detector retrieves the transmitted signals from multiple users from the received symbol vectors. Let us assume a MIMO Base Station (BS) with $Nr$ receive antennas. A total of $Nt$ single antenna users are transmitting a symbol vector $x$ to the BS. The transmitted vector gets distorted by the communication channel $H$ and gets received at the BS as symbol vector $y$. A plethora of algorithms exist for MIMO detection ranging from simple linear detection schemes to highly complex near-optimal or optimal detection schemes. A linear detector uses the pseudo-inverse of the Gramian matrix of $H$ to solve the detection problem. A linear zero-forcing detector can be expressed mathematically as
\begin{equation}\label{eq1}
    x_{\text{hat}} = (H^TH)^{-1}H^Ty
\end{equation}

\begin{algorithm}[t!]
\small
\caption{ABFT-Integrated Newton Iteration}
\label{proposed_algorithm}
\begin{algorithmic}[1]
\Require $H \in \mathbb{C}^{Nr \times N_t}$,\; $y \in \mathbb{C}^{Nr}$,\; noise variance $\sigma^2$, iteration count $\texttt{iter}$,\; tolerance $\epsilon$
\Ensure Estimated signal $\hat{x} \in \mathbb{C}^{N_t}$
\vspace{8pt}
\State \textbf{Real conversion:}\Comment{Gram matrix computation}
 $H_r \gets \begin{bmatrix}\Re\{H\} & -\Im\{H\} \\ \Im\{H\} & \Re\{H\}\end{bmatrix}$,\quad $y_r \gets \begin{bmatrix}\Re\{y\} \\ \Im\{y\}\end{bmatrix}$
\State \textbf{ABFT encoding:} $H_r^{\text{ABFT}} \gets \begin{bmatrix}H_r^\top \\ \mathbf{1}^\top H_r^\top\end{bmatrix}$
\State  $A \gets H_r^{\text{ABFT}} H_r + \sigma^2 \begin{bmatrix}I \\ \mathbf{1}^\top\end{bmatrix}$,\quad $b \gets H_r^{\text{ABFT}} y_r$
\If{$\|A_{\text{chk}} - \sum A_{\text{data}}\| > \epsilon$ or $|b_{\text{chk}} - \sum b_{\text{data}}| > \epsilon$}\\
\Comment{$A_{chk}$ is last row of $A$ and $A_{data}$ is the rest. Same for $b$.}
    \State \textbf{Error observed in preprocessing}
    \State \Return 0
\EndIf 
\vspace{8pt}
\State \textbf{Initialize:} \Comment{MIMO Detection}\\
~~~~~~~~~~~~~$A \gets A_{1:2N_t,1:2N_t}$,
\\ 
~~~~~~~~~~~~~$D \gets 1 ./ \operatorname{diag}(A)$ \\
~~~~~~~~~~~~~$A^{-1}_P \gets [\operatorname{diag}(D),\, D]$ \Comment{Initialize $A_0^{-1}$}\\
~~~~~~~~~~~~~$E \gets 2\begin{bmatrix}I & \mathbf{1}\end{bmatrix}$
\For{$i = 1$ to \texttt{iter}} 
    \State ${A^{-1}}_P \gets {A^{-1}_P}_{(:,1:2N_t)} \cdot (E - A A^{-1}_P)$
    \Comment{ABFT merged into the iterations}
\EndFor
\State \textbf{Solve:} $x \gets \begin{bmatrix} A^{-1}_P(:,1:2N_t) \\ A^{-1}_P(:,2N_t+1)^\top \end{bmatrix} b_{1:2N_t}
$

\If{$|{x_{chk}} - \sum {x_{data}}| > \epsilon$}
\Comment{$x_{chk}$ is last element of $x$ vector and $x_{data}$ is all the rest}
    \State \textbf{Error observed in iterative section}
    \State \Return 0
\EndIf

\State \textbf{Output:} $\hat{x} \gets x_{1:N_t} + j \cdot x_{N_t+1:2N_t}$
\State \Return $\hat{x}$
\end{algorithmic}
\end{algorithm}

In Eq.~\ref{eq1}, the inverse of the Gramian matrix, $A = H^T H$, is multiplied by the matched filter, $b = H^T y$. In a massive MIMO system, which is equipped with a large number of antennas at the base station (BS) to support many single-antenna users, sophisticated and advanced detection mechanisms can be prohibitively complex. Even the inversion of the Gramian matrix, $A$, can introduce very high complexity for linear detection mechanisms, which require costly implementations. Therefore, iterative, approximate inversion-based detectors have become very popular for 5G massive MIMO systems.

We selected the Newton Iteration algorithm to demonstrate the application of Near-Threshold Computing (NTC). In this work, a modified version of the Newton Iteration algorithm was used to incorporate ABFT error detection, as discussed in the next section. The original MI based MIMO algorithm takes the Gramian matrix, $A$, and the matched filter, $b$, as inputs. As the algorithm is iterative, it also takes the number of iterations as an input. The algorithm computes the reciprocals of the diagonal elements of $A$. It then iteratively calculates $A_{\text{inv}}$ using $A$ and the reciprocals of its diagonal elements. Finally, $x_{\text{hat}}$ is computed by multiplying the approximate matrix inverse, $A_{\text{inv}}$, by the matched filter $b$.

\section{Proposed Modified NT MIMO Algorithm }

We propose a low-voltage matrix accelerator with ABFT-based error detection for energy-efficient MIMO computations. This work does not require any specfic HW for ABFT integration \footnote{A HW specific ABFT patented by University of Oulu 2022 \cite{safarpour2021algorithm}.} approach and detect computational errors in MIMO HW/SW simply by inspection of final retrieved signal.

\subsection{Integration of ABFT into MIMO detection algorithm}

The MIMO algorithm with ABFT integrated into it is presented in Algorithm~1. ABFT is utilized in three stages within MIMO detection process, i.e., (i) Gram matrix calculations, (ii) matched filter and (iii) main iterative detection section. Adding ABFT to the \textit{preprocessing stage} during Gram matrix generation, step (i) and matched filter step (ii) is straightforward, as it directly follows the matrix-based ABFT approach introduced earlier. The computational cost of this step depends on how frequently the channel matrix is updated. The essential integration point for ABFT is for step (iii), i.e. the \textit{MIMO detection phase}, where it is embedded within the iterative loop. In this case, minor modifications to the detection algorithm enable ABFT integration with negligible overhead, preserving the original flow and efficiency. At steps 12 to 16 in Algorithm~\ref{proposed_algorithm}, the constant matrices are generated with checksum row/columns depending on the later use. At step 15, the term ${\hat{A}^{-1}}_P=(E-AA^{-1}_P)$ contain ABFT based matrix operation and automatic checksums generation/updates. Later, the previous inverse estimate, denoted with ${\hat{A}^{-1}}_P$ is truncated to remove the checksums, however, the newly generate ${\hat{A}^{-1}}_P$ already is updated with new checksums to preserve checksum property for detecting errors. Finlay, after getting a satisfactory inverse estimate, we solve for $x$ and notice we do not truncate ${\hat{A}^{-1}}_P$ but still need to restructure it slightly to not destroy ABFT's checksum preservation property. Now, $x$ is generated with checksums within as semi-final result. After checking for errors in $x$, in case complex value is preferred, the checksum element is ignored and the data parts of $x$ converted back to real and returned.

\subsection{Computational complexity and ABFT overhead}

The overall Computational Complexity (CC) overhead introduced by ABFT arises from three sources: (i) the computation of the Gram matrix \( A = H^\mathsf{T} H \), (ii) the matched filter output \( b = H^\mathsf{T} y \), and (iii) the iterative MIMO detection process. Among these, the first step is performed during preprocessing and may occur less frequently if the channel matrix \( H \) remains constant over multiple MIMO detections. In such cases, the cost of this step is amortized over many MIMO detection instances. To model this amortization effect, we introduce a coefficient \( 0 < \alpha \leq 1 \), which scales the overhead associated with preprocessing. The parameter \( \alpha \) reflects the relative frequency of Gram matrix and matched filter computations relative to the number of detection instances. When \( \alpha \) is small, preprocessing overhead is negligible; conversely, when \( \alpha \) is close to 1, it dominates.

\subsubsection{Complexity of baseline algorithm}
The computational complexity of Neumann-based detection for a massive MIMO system with \( N_r \) receive antennas and \( N_t \) transmit antennas is expressed in terms of Floating-point Operations (FLOPs). 
The Gram matrix computation \( H^\mathsf{T} H \), using real-valued transformation, requires \(8N_t^2N_r\) FLOPs, while the matched filter output \( H^\mathsf{T} y \) requires \(8N_tN_r\) FLOPs. 
Each iteration of the Newton series involves matrix-matrix multiplications and updates totaling \(16N_t^3\) FLOPs. The final back-substitution step costs an additional \(4N_t^2\) FLOPs. 
Thus, the total computational complexity is:

\[
CC_{\text{MIMO}} = 
\alpha \cdot 8N_t^2N_r + 8N_tN_r + K \cdot 16N_t^3 + 4N_t^2.
\]

\subsubsection{ABFT overheads}
The additional FLOPs introduced by ABFT come from three sources: 
(i) computing input checksums, 
(ii) performing operations on enlarged matrices (due to appended checksum rows), 
and (iii) validating the checksums at the output stage.

In step (i), computing the row checksum of \( H^\mathsf{T} \) costs \(2N_r(2N_t - 1)\) additions. 
The matrix enlargement by one row increases the Gram matrix computation by \(4N_tN_r\) FLOPs, and the checksum validation after multiplication adds another \(4N_tN_r\) FLOPs. In step (ii), the matched filter computation incurs an additional \(2N_r\) FLOPs due to the extra row, while output checksum validation cost is negligible. In step (iii), each Newton iteration has an ABFT-specific overhead of \(6N_t^2 + 4N_t\) FLOPs, as derived from matrix size differences between the augmented and baseline formulations. 
This overhead accumulates over all iterations and becomes relatively small compared to the dominant \(16N_t^3\) FLOPs per iteration for large \( N_t \). Hence,

\[
CC_{\text{overhead}} = 
\alpha \cdot 12N_tN_r + K \cdot (6N_t^2 + 4N_t)
\]

As the problem size grows, the overhead drops very quickly relative to MIMO computations. Assuming $Nr>>Nt$, the overhead simplified to $O(1/N_t)$. Hence, the method is useful for massive MIMO applications that results in large problem formation.

\section{Implementation and Results}

To investigate the proposed approach, first, we investigate the impact of MIMO parameters, such as the number of antennas and users, on interplay of the potential power saving and overheads from adding ABFT. Then, we explore the controlled injection of computational errors through voltage scaling, analyzing its influence on MIMO performance. By addressing these questions, we aim to quantify the trade-offs between energy efficiency and tolerable error levels in MIMO systems, ultimately estimating the achievable power savings through this combined approach.

Our MATLAB fault-injection simulations of MIMO algorithms demonstrate robustness against computational faults. However, without ABFT, one cannot know whether errors have occurred; the MIMO algorithms may tolerate a small fraction of faults, but this tolerance alone cannot be relied upon. As established in prior work~\cite{safarpour2021algorithm, huang1984algorithm, safarpour2021high}, ABFT can detect the majority of errors, with undetected cases being extremely rare. While MIMO remains robust in the presence of a few faults, ABFT provides assurance by explicitly detecting errors rather than depending solely on the algorithm’s inherent resilience.

It was deemed preferable to have the hardware include both a portion, which will be trusted as well as a portion, which will be undervolted and will be allowed to produce errors during testing. For this reason, the MIMO detection application was divided between a software part and a hardware-accelerated part. The hardware-accelerated part is represented by a matrix operation unit capable of matrix multiplication, matrix addition and matrix subtraction operations. These operations are the bulk of the computations of the MIMO detection algorithm, using Newton's iterative method. The outputs of these matrix operations can also be protected using ABFT. Other aspects of the program are handled on the software side of the test platform. Having a matrix accelerator unit with fixed-size inputs (in our case 16x16) requires padding small matrices with zeros to fit the accelerator's input dimensions as well as dividing large matrices to smaller sub-matrices. This feature of the hardware design has implications when it comes to the timing overheads, introduced by adding ABFT to the program. 

The test hardware aims to represent a situation in the field, where the MIMO algorithm is partly accelerated with hardware, which cannot be modified without expensive re-fabrication, but where the parts of code running on a general-purpose processor could be changed with a software update.

\begin{figure}[h]
  \begin{center}
  \includegraphics[width=2.8in]{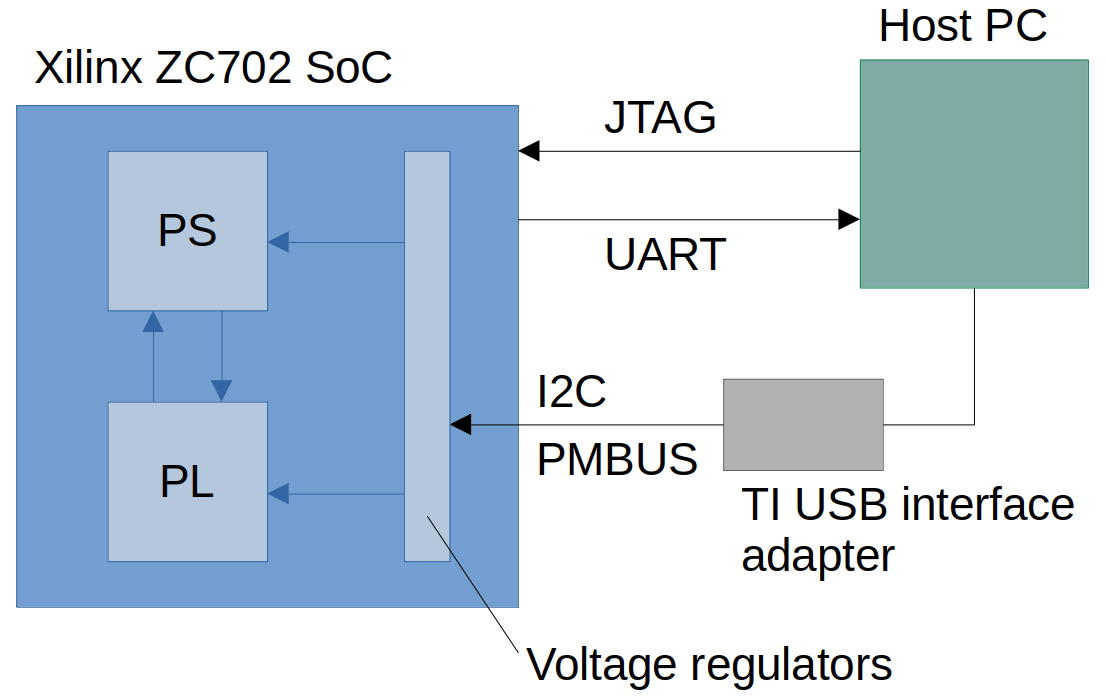}
  \vspace{-5pt}
  \caption{Zynq SoC setup for lower voltage experiments.}\label{setup}
  \end{center}
    \vspace{-5pt}
\end{figure}

The hardware design was implemented on a Xilinx Zynq ZC702 SoC, featuring both Programmable Logic (PL) as well as a dual-core ARM processor (PS). The PL was configured to house the matrix accelerator unit. The creation of simulated input data, control logic and modulation was done on the ARM core of the board. Only the PL side of the board was undervolted and was at risk of producing errors during testing. The design for the matrix accelerator was written in C and synthesized using Xilinx Vivado HLS. The FPGA board design was done on Xilinx Vivado and the accompanying PS software was written and tested on Xilinx SDK\footnote{Source code is available online at \url{https://github.com/NortHund/NTV_FT_MIMO}.}. Voltage levels were set on the board using Texas Instruments Fusion Digital Power Designer and power readings were taken in the test program itself, using PMBUS commands. A diagram of the test hardware setup is shown in figure \ref{setup}. All tests were run with 3 iterations on the
algorithm.


\subsection{Overheads}


The sizes of the large matrices used in MIMO detection are determined by the amount of antennas on the transmitting and the receiving side. As the accelerator accepts inputs of two fixed-size matrices, the large matrices are divided into smaller sub-matrices that can be streamed to the accelerator.

As $N_t$ and $N_r$ increase, and their ratio grows, the relative overhead introduced by ABFT decreases due to its inverse relation with matrix size. An exception arises when the matrix dimensions exactly match the size of our matrix unit. In this boundary case, the matrix cannot be mapped efficiently onto the acceleration unit, leading to additional operations, primarily in the form of zero-padded matrix multiplications. These extra computations increase the measured overhead. Nonetheless, for the general case across a wide range of matrix dimensions, the observed overhead remains relatively small, typically within $3\%$ to $7\%$ of the execution time. The overheads were measured by taking the times of running MIMO detection algorithm 100 times on hardware, with and without ABFT and comparing the results.


\subsection{Undervolting experiments}
\begin{figure}[h]
  \begin{center}
  \includegraphics[width=3.6in]{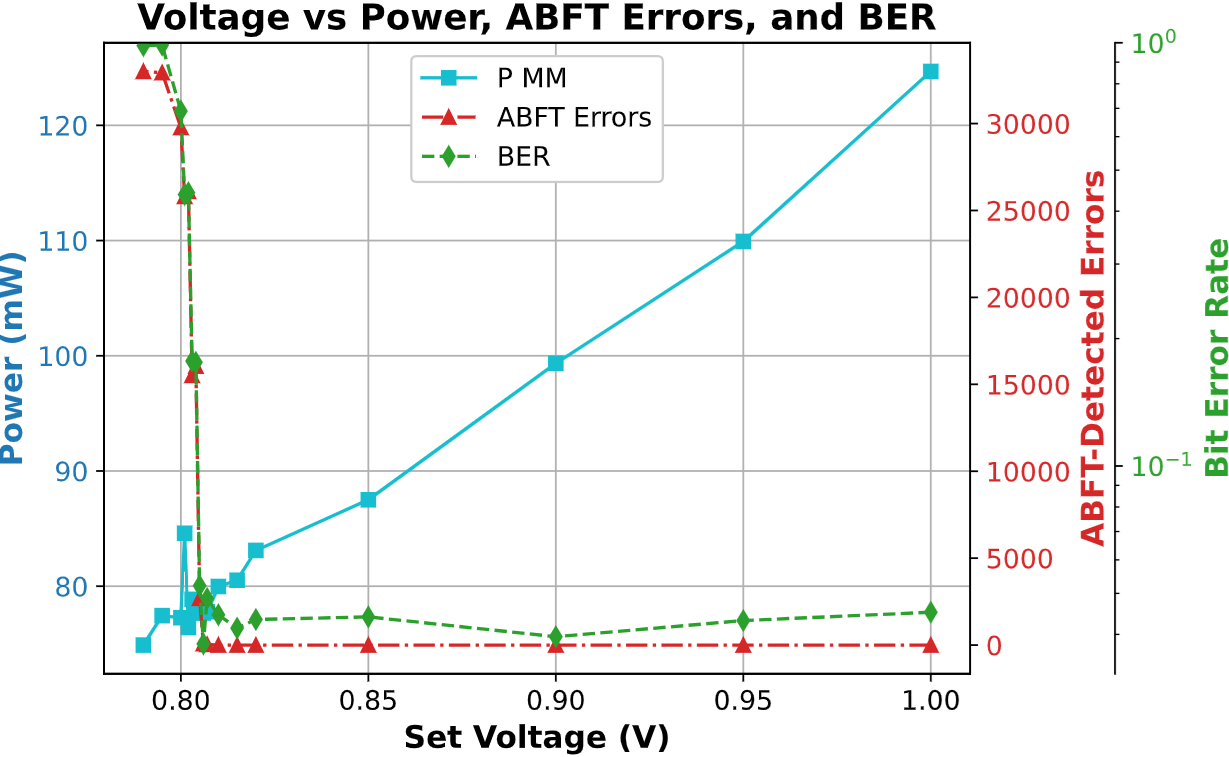}
  \vspace{-10pt}
  \caption{Operating voltage and amount of detected ABFT errors, bit error rate and power usage}\label{undervolt}
  \end{center}
    \vspace{-5pt}
\end{figure}

\vspace{-0.03cm}
Tests were run while lowering the operating voltage of the PL side of the board, to see the effect on power usage and detection accuracy. The results are shown in figure \ref{undervolt} for the default frequency of 100 MHz.

The power usage was measured from within the test program, while running a loop of matrix multiplication operations, after the two input matrices were streamed in to the accelerator, but before the program starts waiting for the output. The average power reading from 1000 passes was taken for each voltage. As the test program was single-threaded, measuring the power usage this way doesn't give an accurate representation on the total average energy use of the MIMO detection algorithm, with both the PS and PL side taken into account. Instead, this method tests only the PL side under load, which also is the only side being undervolted. As was expected, lowering the operating voltage on the PL side quickly reduced the power usage of the chip. Comparing the power usages at the default voltage of 1.0 $V$ and the voltage at Point of First Failure (PoFF), which was 0.807 $V$, a 36\% reduction in power use was observed. 

The ABFT-detected errors were measured by running the input with $N_t=8$ and $N_r=64$ for a thousand cycles, while adding together all checksum mismatches reported by the program. As the test program uses floating point numbers, ABFT cannot discern very small errors in the output from quantization noise. It is also possible that the error happens while calculating the ABFT checksums, not in the MIMO calculations. These situations produce false negative and false positive detections respectively. A false positive detection is less of an issue, as it only causes a small timing delay, when the faulty output has to be re-run. False negative detections on the other hand could lead to incorrect values propagating forward in the system. However, the MIMO algorithm is inherently tolerant towards very small errors in the output. During testing false negatives were detected by comparing the hardware-accelerated output to a reference output calculated completely in PS. Some false negatives were detected at around PoFF voltage, but they never resulted in an impact to the bit error rate.

Examining the bit error rate of the detection algorithm for each operating voltage point reveals also something interesting. The tests were run with a Signal-to-Noise-Ratio (SNR) of $10$, $Nt=8$ and $Nr=64$ for a thousand cycles, taking the average of the outputs. For the error-free voltage range, the Bit Error Rate (BER) stays at ~4\%. Then quickly after the PoFF, at 0.8 $V$, the bit error rate shoots up to almost 70\%. The experiments suggest that the accuracy of MIMO is not affected by very small errors in the computations, but as the errors grow larger, they both affect the bit error rate greatly and are easier to detect with ABFT. At 0.8 $V$, no false negatives were detected any more. Thus, the inherent robustness of MIMO should not be solely relied upon and ABFT detections used to verify the results.

Additional tests were ran after lowering the clock frequency of PL, to see how far the voltage could be reduced on lower speeds. The results are displayed in Table \ref{tab:underclock}. The point of first failure was tested by running a hundred passes of the MIMO detection algorithm with $N_r=64$, $N_t=8$ with 3 iterations, while seeing if ABFT reports any errors. Then the power usage was measured by taking the average after a thousand passes of matrix multiplication operations.

\begin{table}[t!]
\footnotesize
\centering
\caption{Power consumption of MIMO matrix accelerator at different clock frequencies and voltages.}
\vspace{-5pt}
\label{tab:power_savings}
\renewcommand{\arraystretch}{1.3} 
\setlength{\tabcolsep}{4pt} 
\begin{tabular}{lcccc}
\toprule
\textbf{PL clock frequency} & \textbf{100 MHz} & \textbf{75 MHz} & \textbf{50 MHz} & \textbf{25 MHz} \\
\midrule
\textbf{$V_{PoFF} (mV)$} & $807$ & $765$ & $695$ & $633$ \\
\textbf{$V_{crash} (mV)$} & $730$ & $680$ & $670$ & $620$ \\
\textbf{$P$($mW$) @$V_{Default}$} & $119$ & $117$ & $102$ & $75$ \\
\textbf{$P$($mW$) @$V_{PoFF}$} &$76$ & $59 $ & $32$ & $18$\\
\bottomrule
\end{tabular}
\label{tab:underclock}
\end{table}




\section{Discussion}

Undervolting the processor while managing computational errors through ABFT provides an interesting way of saving energy, while accepting only small penalties in terms of performance loss. The biggest limitation of the method is in how it can be efficiently applied only to such use-cases in which most of the computations are large-scale matrix operations. One such use-case is MIMO detection.
The approach proposed in this research uses a single matrix accelerator unit for both the reference implementation of MIMO detection as well as the ABFT-equipped variant. All necessary changes for adding ABFT were done in software, on the PS side of the Zynq board. The benefit of taking this path is that ABFT can be applied elegantly to the entire MIMO detection algorithm, resulting in low computational overheads - both theoretical and observed. Still, with specific amounts of users communicating simultaneously with the tower, the overhead spikes up, owing to the fact that the large operations need to be divided into smaller sub-matrix operations. Another way would have been to have a single ABFT-enabled matrix accelerator, with fault tolerance being synthesized to be a part of the hardware. This would have resulted in a consistent overhead, although mathematically it would have been higher than in the typical cases for the proposed approach. ABFT in matrix multiplication hardware is also well-researched as it is.
Yet another way would have been to add ABFT to the state-of-the-art generation of dedicated MIMO accelerators in hardware \cite{aliagha2025scissors}. This could be an interesting topic of research for the future, although the results of it would not be as applicable to the current state of MIMO detection accelerators post-fabrication as our proposed approach hopefully is. Further research is also needed for experimenting with different frequencies, e.g. overclocking the circuit and pushing the voltage to the thermal limits, e.g., exploiting Temperature Effect Inversion \cite{lee2014dynamic}.






\vspace{-3pt}
\section{Conclusion}

Using undervolting in conjunction with ABFT to protect against timing errors provides an effective way of reducing power usage in large-scale MIMO systems. With the configuration tested, an energy saving potential of ~36\% was observed, while observing a small loss in throughput - around ~3\% in a typical situation. Lowering the operating voltage of the system to the point of first failure and beyond quickly increased the amount of computational errors and worsened the detection accuracy of the system. However, adding ABFT to Newton's Iterative method protected the output fully, and with no impact on the bit error rate from timing errors due to low voltage operation. 


\section{Acknowledgment}

This work was supported by 6G Flagship (Grant Number 369116) funded by the Research Council of Finland and the Finland’s Ministry of Education and Culture MIcroELectronics (MIELi) doctoral school pilot program.

\ifCLASSOPTIONcaptionsoff
  \newpage
\fi

\bibliographystyle{IEEEtran}


\vfill

\end{document}